\documentclass[
    ,final            % use final for the camera ready runs
  ]
  {aipproc}

\layoutstyle{8x11single}

\begin{document}

\title{Recurrence plots and chaotic motion around Kerr black hole}


\author{Ond\v{r}ej Kop\'{a}\v{c}ek}{
  address={Astronomical Institute, Academy of Sciences, Prague, Czech Republic}
}

\author{Ji\v{r}\'{i} Kov\'{a}\v{r}}{
  address={Institute of Physics, Silesian University, Opava, Czech Republic}
}

\author{Vladim\'{\i}r Karas}{
  address={Astronomical Institute, Academy of Sciences, Prague, Czech Republic}
}

\author{Zden\v{e}k Stuchl\'{i}k}{
 address={Institute of Physics, Silesian University, Opava, Czech Republic}
}

\begin{abstract}
We study the motion of charged test particles around a
Kerr black hole immersed in the asymptotically uniform magnetic field,
concluding that off-equatorial stable orbits are allowed in this
system. Being interested in dynamical properties of these astrophysically relevant orbits we employ rather novel approach
based on the analysis of recurrences of the system to the vicinity
of its previous states. We use recurrence plots (RPs) as a tool to
visualize recurrences of the trajectory in the phase space.
Construction of RPs is simple and straightforward regardless of the
dimension of the phase space, which is a major advantage of this
approach when compared to the ``traditional'' methods of the numerical
analysis of dynamical systems (for instance the visual survey of
Poincar\'{e} surfaces of section, evaluation of the Lyapunov spectra
etc.). We show that RPs and their quantitative measures (obtained
from recurrence quantification analysis -- RQA) are powerful tools
to detect dynamical regime of motion (regular vs. chaotic) and
precisely locate the transitions between these regimes.
\end{abstract}

\maketitle

%%%%%%%%%%%%%%%%%%%%%%%%%%%%%%%%%%%%%%%%%%%%
%% MAINMATTER
%%%%%%%%%%%%%%%%%%%%%%%%%%%%%%%%%%%%%%%%%%%%

\section{Introduction}
In this work we continue our effort \cite{halo3,halo2,kovar08} to understand the dynamic properties
of charged test particles being exposed to the electromagnetic test
fields surrounding compact objects -- neutron stars and black holes.
As we bear astrophysical motivation in our mind we choose such
fields and such background geometries which combine themselves in reasonable
models of real situations occurring in the vicinity of these
objects. Survey of the test particle trajectories might be regarded
as the one particle approximation to the complex dynamics of the
astrophysical plasma.

In this contribution we investigate the motion of charged test particles
around the Kerr black hole immersed into the asymptotically uniform
magnetic field (Wald's solution \cite{wald}).  This field allows
motion in the off-equatorial lobes if the parameters of the system are
chosen carefully \cite{halo2}. By off-equatorial lobes we mean closed equipotential surfaces which enclose the stable off-equatorial circular orbit (so called halo orbit). They symmetrically intersect poloidal $(r,\theta)$ plane as seen in Fig. \ref{traj}. We investigate motion of the charged particles in these lobes. We are
particularly curious what is the dynamic regime of motion (chaotic
versus regular) and how does it change if we alter some of the
parameters. Besides the standard technique of Poincar\'{e} surfaces of
section we employ recurrence analysis \cite{marwan} and show that
recurrence plots might be regarded as an alternative tool to surfaces
of section. Quantitative analysis of recurrences appears to be very
useful tool to locate the transitions between these regimes in the terms
of selected control parameter.

\section{Kerr black hole in asymptotically uniform magnetic field}
The Kerr metric in Boyer-Lindquist coordinates $t,\:r, \:\theta,\:\phi$
is given as follows \cite{mtw}:
\begin{equation}
\label{kn}
ds^2=-\frac{\Delta}{\Sigma}\:[dt-a\sin{\theta}\,d\phi]^2+\frac{\sin^2{\theta}}{\Sigma}\:[(r^2+a^2)d\phi-a\,dt]^2+\frac{\Sigma}{\Delta}\;dr^2+\Sigma
d\theta^2,
\end{equation}
where we set
\begin{equation}
\label{knleg} {\Delta}\equiv{}r^2-2Mr+a^2,\;\;\;
\Sigma\equiv{}r^2+a^2\cos^2\theta,
\end{equation}
and $M$ stands for the mass of the black hole and $a$ for its
specific angular momentum.

In order to incorporate the large scale magnetic field to our
considerations we employ the Wald's test field solution \cite{wald}.
Vector potential of this axisymmetric test field may be expressed in
the terms of Kerr metric coefficients as follows:
\begin{equation}
\label{waldpot1}
A_t=\frac{1}{2}B_{0}\left(g_{t\phi}+2a\,g_{tt}\right)-\frac{\tilde{Q}}{2}g_{tt}-\frac{\tilde{Q}}{2},
\end{equation}

\begin{equation}
\label{waldpot2}
A_{\phi}=\frac{1}{2}B_{0}\left(g_{\phi\phi}+{2a}g_{t\phi}\right)-\frac{\tilde{Q}}{2}g_{t\phi},
\end{equation}

where $\tilde{Q}=Q/M$ stands for the specific test charge of the black hole
(unlike the Kerr-Newman solution it does not enter the metric).
The terms containing $\tilde{Q}$ in above equations may thus be
identified with the components of the vector potential of the
Kerr-Newman solution \cite{mtw}. Asymptotic behavior of the components
justifies the identification of the parameter $B_0$ with the
strength of the originally uniform magnetic field into which the
Kerr black hole has been immersed. Wald \cite{wald} has shown that
in the case of parallel orientation of the spin and the magnetic
field $B_{0}$ the black hole selectively accretes positive charges
(negative for the antiparallel orientation) until it is charged to
the value $\tilde{Q}_{\rm{W}}=2B_{0}a$. Although the use of the Wald's
charge $\tilde{Q}_{\rm{W}}$ is not obligatory (since the presence of
another charging mechanism could be considered for the astrophysical
black holes) it may be regarded as a preferred value.

\section{Recurrence plots and recurrence quantification analysis}
Besides various ``traditional'' methods of the numerical analysis of
dynamical systems (for instance the visual survey of Poincar\'e surfaces
of section,  evaluation of the Lyapunov spectra \cite{lce} etc.) it
appears useful to employ rather novel approach based on the analysis
of recurrences of the system to the vicinity of its previous states.
Recurrence plots (RPs) as a tool to visualize recurrences of the
trajectory in the phase space were introduced by Eckmann et al in
1987 \cite{eckmann}. RP method is based on the examination of the binary
values that are constructed from the phase space trajectory $\vec{x}(t)$.
Construction of RPs is simple and straightforward regardless of the
dimension of the phase space which is a major advantage of this
approach. Binary values of the recurrence matrix $\mathbf{R}_{ij}$
may be formally expressed as follows:

\begin{equation}
\label{rpdef}
\mathbf{R}_{ij}(\varepsilon)=\Theta(\varepsilon-||\vec{x}(i)-\vec{x}(j)||),\;\;\;
i,j=1,...,N
\end{equation}
where $\varepsilon$ is a predefined threshold parameter, $\Theta$
represents Heaviside step function and $N$ specifies the sampling
frequency which is applied to the examined time period of the
trajectory $\vec{x}(t)$.

Selection of the norm $||.||$ which should be used to detect
recurrences in the phase space is not straightforward. Although
simple norms like $L^2$ (Euclidean norm) are usually applied
directly in this context, we want to reflect the curvature of the
spacetime as much as possible. Thus we measure distances in the
ZAMO's hypersurfaces of simultaneity following standard $3+1$
splitting procedure \cite{macdonald}. Projection tensor we apply to the
phase space constituents $x^{\mu}$, $\pi_{\mu}$ is given as follows:
\begin{equation}
\label{3Dmetric} \gamma_{\mu\nu}=g_{\mu\nu}+u_{\mu}u_{\nu},
\end{equation}
where $g_{\mu\nu}$
is a spacetime metric (\ref{kn}) and $u_{\mu}$ stands for the coordinate components of ZAMO's four-velocity $u$ which may be expressed as follows \cite{bardeen}:
\begin{equation}
\label{ZAMOvel}
u=\frac{\sqrt{A}}{\sqrt{\Delta\Sigma}}\left(\frac{\partial}{\partial{}t}+\Omega\frac{\partial}{\partial\phi}\right),
\end{equation}
setting $A\equiv(r^2+a^2)^2-a^2\Delta\sin^2\theta$ and $\Omega=\frac{2a}{A}Mr$.

Consequent order estimates lead to the conclusion that the
computation of the recurrence matrix $\mathbf{R}_{ij}$ may be
carried out using Euclidean norm applied to the following set of
coordinates $(\sqrt{g_{rr}}r,
\sqrt{g_{\theta\theta}}\theta,\sqrt{g_{\phi\phi}}\phi,\sqrt{g^{rr}}\pi_{r},\sqrt{g^{\theta\theta}}\pi_{\theta},\frac{1}{\sqrt{g_{\phi\phi}}}L)$,
provided that the threshold $\varepsilon$ as the upper limit of the distance at which we shall apply this
norm is significantly smaller than the characteristic scale of the
spacetime curvature. As a length scale of curvature we may consider $P=K^{-1/4}$, where $K=R^{\mu\nu\sigma\epsilon}R_{\mu\nu\sigma\epsilon}$ represents the Kretschmann scalar evaluated by contracting Riemann curvature tensor. We must numerically check whether the assumption  $\frac{\varepsilon^2}{P^2}\ll{}1$ holds along a given trajectory.

Binary valued matrix $\mathbf{R}_{ij}$ represents the RP which we
get by assigning a black dot where $\mathbf{R}_{ij}=1$ and leaving a
white dot where $R_{ij}=0$. Both axis represent the time period over
which the data set (phase space vector) is being examined. RP is
thus symmetric and the main diagonal is always occupied by the line
of identity (LOI).

Structures in the RP encode surprisingly large amount of information
about the dynamics of the system \cite{thiel}. For the purpose of
deciding whether the particular trajectory is regular or chaotic the
determining factor is a presence and amount of diagonal structures
in the RP. Diagonal lines in the RP reflect the time segments of the
phase space trajectory where the system evolves similarly. It
captures the epoch when the trajectory runs almost parallel to its
previous segment i.e. runs inside the $\varepsilon$--tube around
this segment \cite{marwan}. Hence integrable systems with regular dynamics result in strongly diagonally oriented RPs. On the
other hand when the motion is chaotic the diagonal lines are rather
short (because the trajectory tends to diverge quickly) and more
complicated structures are found in the RP.

The recurrence quantification analysis (RQA) \cite{marwan} takes number
of statistic measures of the recurrence matrix $\mathbf{R}_{ij}$. First
of all we define the recurrence rate $RR$ as a density of points in the RP:
\begin{equation}
\label{RR}
 RR(\varepsilon)\equiv\frac{1}{N^2}\sum_{i,j=1}^N\mathbf{R}_{i,j}(\varepsilon).
\end{equation}

Then we turn our attention to the diagonal segments in the RP whose length
basically draws distinction between regularity and chaos. Histogram
$P(\varepsilon,l)$ recording the number of diagonal lines of the length
$l$ is formally given as follows:
\begin{equation}
\label{diaghist}
P(\varepsilon,l)=\sum^N_{i,j=1}(1-\mathbf{R}_{i-1,j-1}(\varepsilon))(1-\mathbf{R}_{i+l,j+l}(\varepsilon))\prod_{k=0}^{l-1}\mathbf{R}_{i+k,j+k}(\varepsilon).
\end{equation}
Histogram $P(\varepsilon,l)$ is used to define the determinism $DET$ as
a fraction of the recurrence points which form diagonal lines of length
at least $l_{\rm{min}}$ to all recurrence points:
\begin{equation}
\label{det}
DET\equiv\frac{\sum^N_{l=l_{\rm{min}}}lP(\varepsilon,l)}{\sum^N_{l=1}lP(\varepsilon,l)},
\end{equation}
average length of the diagonal line $L$ (where only lines of length at
least $l_{\rm{min}}$ count):
\begin{equation}
\label{Lav}
L\equiv\frac{\sum^N_{l=l_{\rm{min}}}lP(\varepsilon,l)}{\sum^N_{l=l_{\rm{min}}}P(\varepsilon,l)},
\end{equation}
and divergence $DIV$ as an inverse of the length of the longest
diagonal line $L_{\rm{max}}$:
\begin{equation}
\label{div}
DIV\equiv\frac{1}{L_{\rm{max}}}.
\end{equation}
Since $DIV$ is in its very nature closely related to the divergent
features of the phase space trajectory it was originally
\cite{eckmann} claimed to be directly related to the largest
positive Lyapunov characteristic exponent $\lambda_{\rm{max}}$.
Nevertheless theoretical considerations justify the use of $DIV$ as
an estimator only for the lower limit of the sum of the positive
Lyapunov exponents \cite{marwan}. On the other hand strong
correlation between $DIV$ and $\lambda_{\rm{max}}$ still arises in
numerical experiments \cite{trulla}.

It appears useful to perform the analogous statistics also for the vertical
(horizontal respectively, since the RP is symmetric with respect to LOI) segments in RPs which are generally connected with
periods in which the system is slowly evolving (laminar states). To
this end histogram $P(\varepsilon,v)$ recording the number of
vertical lines of length $v$ is constructed as follows:
\begin{equation}
\label{verthist}
P(\varepsilon,v)=\sum^N_{i,j=1}(1-\mathbf{R}_{i,j}(\varepsilon))(1-\mathbf{R}_{i,j+v}(\varepsilon))\prod_{k=0}^{v-1}\mathbf{R}_{i,j+k}(\varepsilon).
\end{equation}

In analogy with the diagonal statistics histogram $P(\varepsilon,v)$ is
used to define vertical RQA measures. Laminarity $LAM$ is defined as
a fraction of recurrence points which form vertical lines of length
at least $v_{\rm{min}}$ to all recurrence points:
\begin{equation}
\label{lam}
LAM\equiv\frac{\sum^N_{v=v_{\rm{min}}}vP(\varepsilon,v)}{\sum^N_{v=1}vP(\varepsilon,v)},
\end{equation}
trapping time $TT$ is an average length of the vertical line (where only
lines of the length at least $v_{\rm{min}}$ count):
\begin{equation}
\label{tt}
TT\equiv\frac{\sum^N_{v=v_{\rm{min}}}vP(\varepsilon,v)}{\sum^N_{v=v_{\rm{min}}}P(\varepsilon,v)}
\end{equation}
and finally also the length of the longest vertical line
$V_{\rm{max}}$ might be considered as an analogue of the diagonal
$L_{\rm{max}}$.

\section{Test particle motion}
\subsection{Equations of motion}
Generalized Hamiltonian (``super Hamiltonian``) which characterizes the
dynamics of the test particle of charge $q$ and mass $m$ is given as
follows \cite{mtw}:
\begin{equation}
\label{SuperHamiltonian}
\mathcal{H}=\frac{1}{2}g^{\mu\nu}(\pi_{\mu}-qA_{\mu})(\pi_{\nu}-qA_{\nu}),
\end{equation}
where $\pi_{\mu}$ is the generalized (canonical) momentum and $A_{\mu}$ denotes the vector potential related to the electromagnetic field tensor as $F_{\mu\nu}=A_{\nu,\mu}-A_{\mu,\nu}$.

\begin{figure}[htb]
\centering\label{wald_abc}
\includegraphics[scale=.95,clip]{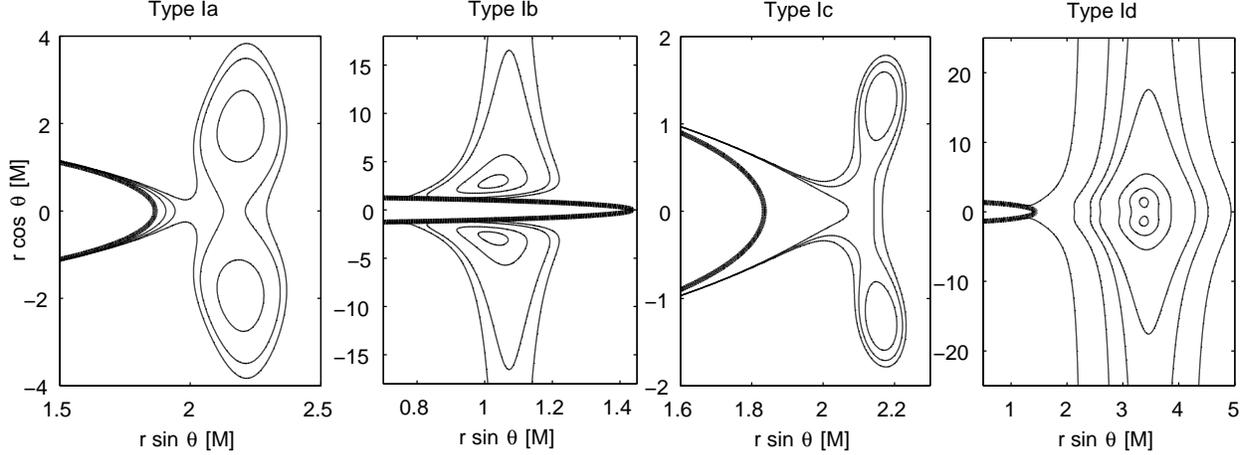}
\caption{Classification of possible topologies of off-equatorial potential lobes appearing above the event horizon (bold line) of the Kerr black hole immersed into the Wald's test field. See \cite{halo2} for details.}
\end{figure}

The Hamiltonian equations of motion are given in a standard way:
\begin{equation}
\label{HamiltonsEquations}
{\rm d}x^{\mu}/{\rm d}\lambda=\partial
\mathcal{H}/\partial \pi_{\mu},\;\;\;\;\; {\rm d}\pi_{\mu}/{\rm d}\lambda=-\partial\mathcal{H}/\partial x^{\mu},
\end{equation}
where $\lambda=\tau/m$ denotes the affine parameter and $\tau$ the proper
time.

In the case of stationary and axially symmetric systems we
immediately identify two constants of motion from the second
Hamiltonian equation, namely energy $E$ and angular momentum $L$:
\begin{equation}
\label{Momenta} \pi_t=p_t+qA_t\equiv-E,\;\;\;\;
\pi_{\phi}=p_{\phi}+qA_{\phi}\equiv L.
\end{equation}

\subsection{Effective potential}
We depart from the normalization of the four-momentum
\begin{equation}
\label{normalizace}
g^{\mu\nu}p_{\mu}p_{\nu}=g^{\mu\nu}(\pi_{\mu}-qA_{\mu})(\pi_{\nu}-qA_{\nu})=-m^2,
\end{equation}
 which is conserved along the Hamiltonian flow in autonomous systems. In stationary and axially symmetric situation it allows to express the condition for simultaneous turning point in $r$, $\theta$ coordinates, i.e. simultaneous zero points of $u^r=p^r/m=g^{rr}\pi_{r}/m$ and $u^{\theta}=p^{\theta}/m=g^{\theta\theta}\pi_{\theta}/m$. These points connect in curves which represent natural boundary for the test particle motion at given energetic level. This two-dimensional (i.e. related to the motion in two coordinates) effective potential describing the motion of the charged test particle takes the following form:

\begin{equation}
\label{effpot} V_{\rm
eff}(r,\theta;\;a,\:\tilde{q}\tilde{Q},\:\tilde{q}B_{0},\: \tilde{L})=\frac{-\beta+\sqrt{\beta^2-4\alpha\gamma}}{2\alpha},
\end{equation}
denoting
\begin{equation}
\label{alfa}
 \alpha=-g^{tt},
\end{equation}
\begin{equation}
\label{beta}
\beta=2[g^{t\phi}(\tilde{L}-\tilde{q}A_{\phi})-g^{tt}\tilde{q}A_{t}],
\end{equation}
\begin{equation}
\label{gama}
\gamma=-g^{\phi\phi}(\tilde{L}-\tilde{q}A_{\phi})^2-g^{tt}\tilde{q}^2A_t^2+2g^{t\phi}\tilde{q}A_t(\tilde{L}-\tilde{q}A_{\phi})-1,
\end{equation}

where we introduce "specific'' quantities $\tilde{L}\equiv{}L/m$ and
$\tilde{q}\equiv{}q/m$. Setting the specific energy $\tilde{E}\equiv{}E/m$ of the given test particle we obtain isocontours forming the boundary of allowed region which is accessible to its trajectory. Since the quantities  $\tilde{q}$, $\tilde{Q}$ and $B_{0}$ appear only in terms $\tilde{q}\tilde{Q}$ and $\tilde{q}B_{0}$ we only need to specify values of these two products to uniquely determine the test particle (provided that remaining parameters are already set).

\begin{figure}[htb]
\centering\label{traj}
\includegraphics[scale=.85,clip]{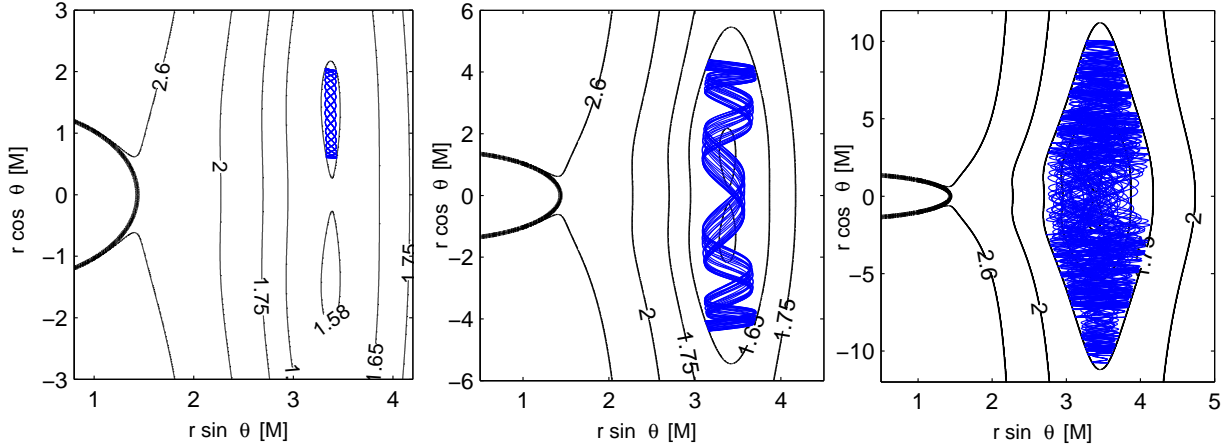}
\caption{The test particle ($\tilde{L} =6M$, $\tilde{q}B_{0}=M^{-1}$ and $\tilde{q}\tilde{Q}=1$) is
launched from the locus of the off-equatorial potential minima $r(0)=3.68\;M$, $\theta(0)=1.18$ with $u^r(0)=0$ and various values of the energy $\tilde{E}$. In the left panel we set $\tilde{E}=1.58$ and we observe ordered off-equatorial motion. For the energy of $\tilde{E}=1.65$ cross-equatorial regular motion is observed (middle panel). The trajectory occupies only a part of allowed potential lobe, regardless the length of the
integration period. Finally in the right panel with $\tilde{E}=1.75$ we observe irregular motion whose trajectory  would ergodically fill whole allowed region after the sufficiently long integration time. We show that the motion
is chaotic in this case. Spin of the black hole is $a=0.9\:M$ and its event horizon is depicted by the bold line.}
\end{figure}

In order to locate the halo orbit and related off-equatorial potential lobe we search for the local minima of the potential $V_{\rm eff}$. Direct approach based on the investigation of the conditions $\frac{\partial{}V_{\rm eff}}{\partial{}r}=0$, $\frac{\partial{}V_{\rm eff}}{\partial\theta}=0$ and the determinant of the Hessian matrix being positive, leads to the set of very intricate algebraic equations. However, when we reformulate these conditions using the force formalism we obtain cubic equation which we are able to solve in general. See \cite{kovar08} for details. We found \cite{halo2}
that potential minima may appear above the event horizon of the black hole, i.e. that off-equatorial stable orbits are allowed in this setup -- see Fig.~\ref{wald_abc}. Besides the
off-equatorial lobes the potential may form another remarkable
structure -- endless potential valley of almost constant depth which
runs parallelly to the symmetry axis. Thus the particle may escape
to infinity from the equatorial plane if its parameters are suitably
chosen. Such orbits may form highly collimated jet-like structure and may also provide a charge separation mechanism as was recently reported in \cite{preti}. Motion in the potential valleys we further discuss in \cite{halo3}. Here we concern ourselves with the motion in the closed potential lobes only.

\begin{figure}[htb]
\centering\label{kntraj}
\includegraphics[scale=0.43, clip]{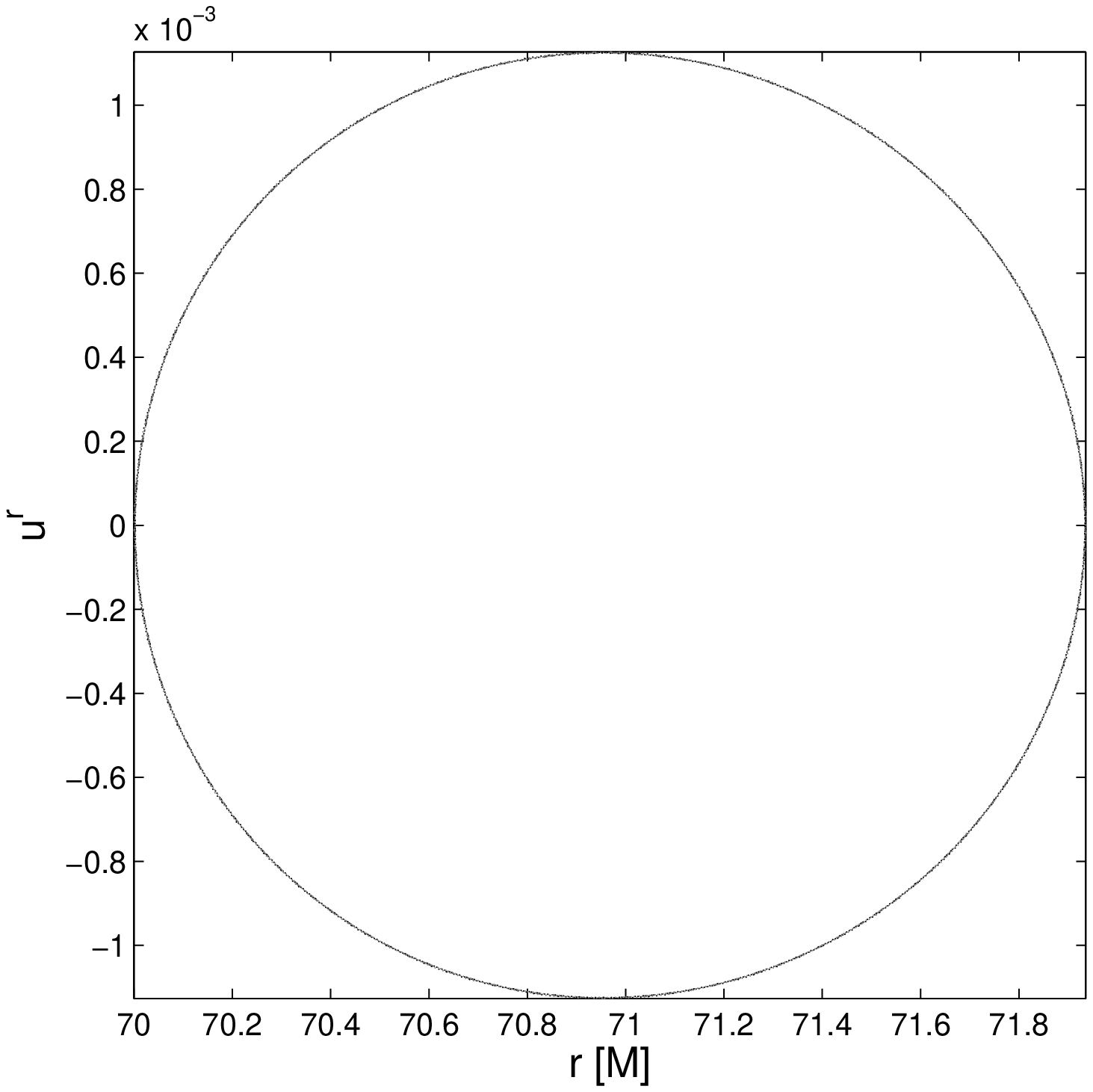}
\includegraphics[scale=0.38, clip]{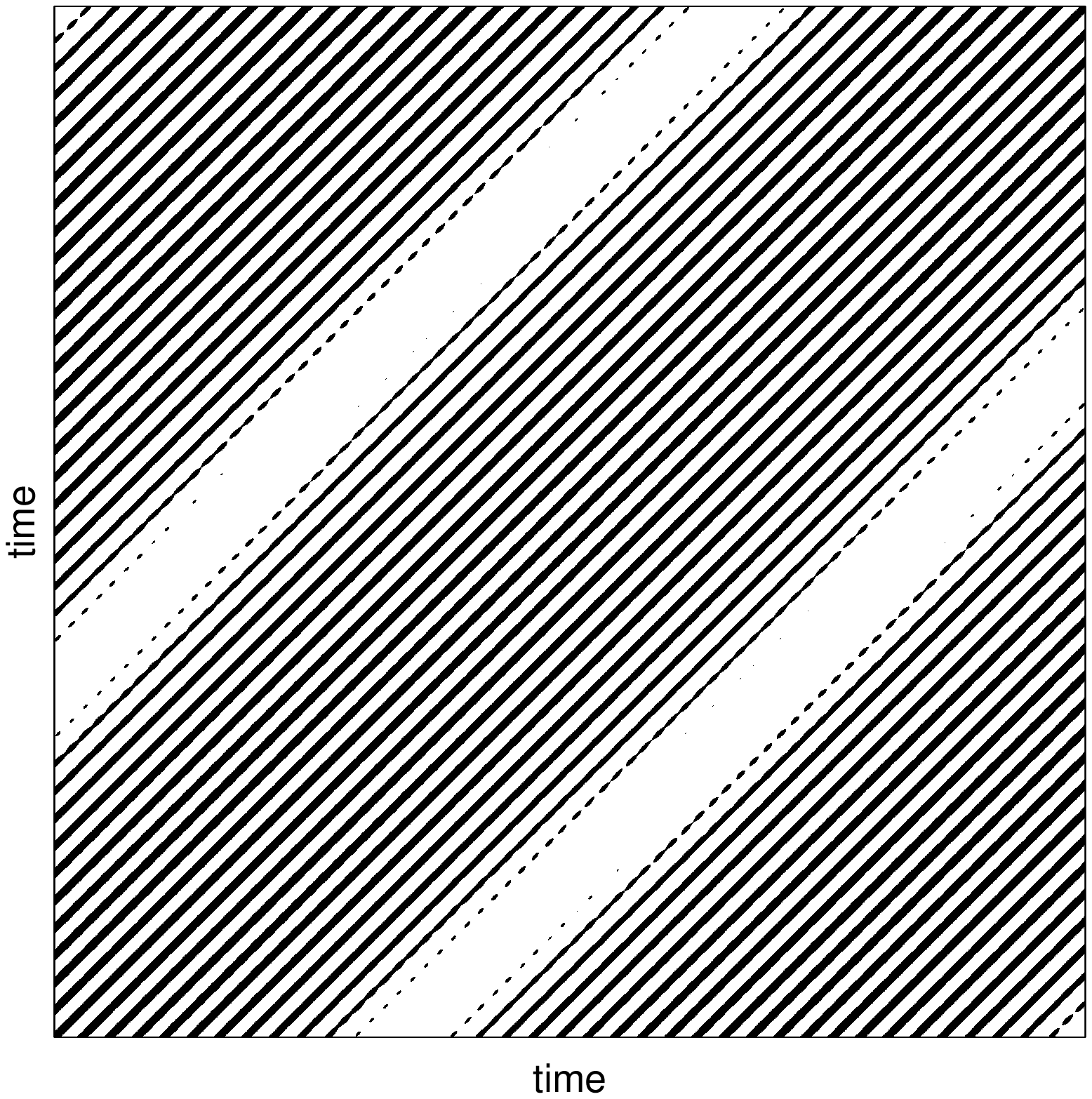}
\caption{For the sake of comparison we present the surface of section
and recurrence plot for a regular trajectory in the fully integrable
system of charged test particle ($\tilde{E}=0.9965$, $\tilde{L} =6M$,
$\tilde{q}=10^{17}$, $r(0)=70\:M$, $\theta(0)=\theta_{\rm section}=\pi/2$ and $u^r(0)=0$) in the pure Kerr-Newman spacetime
($\tilde{Q}=5\times10^{-18}$, $a=0.9\:M$) endowed with the fourth
Carter's constant of motion $\mathcal{L}$ \cite{mtw}. Long diagonals
parallel to the LOI  are a general hallmark of regularity in the RPs.}
\end{figure}

\begin{figure}[htb]
\centering \label{waldreg}
\includegraphics[scale=.43,clip]{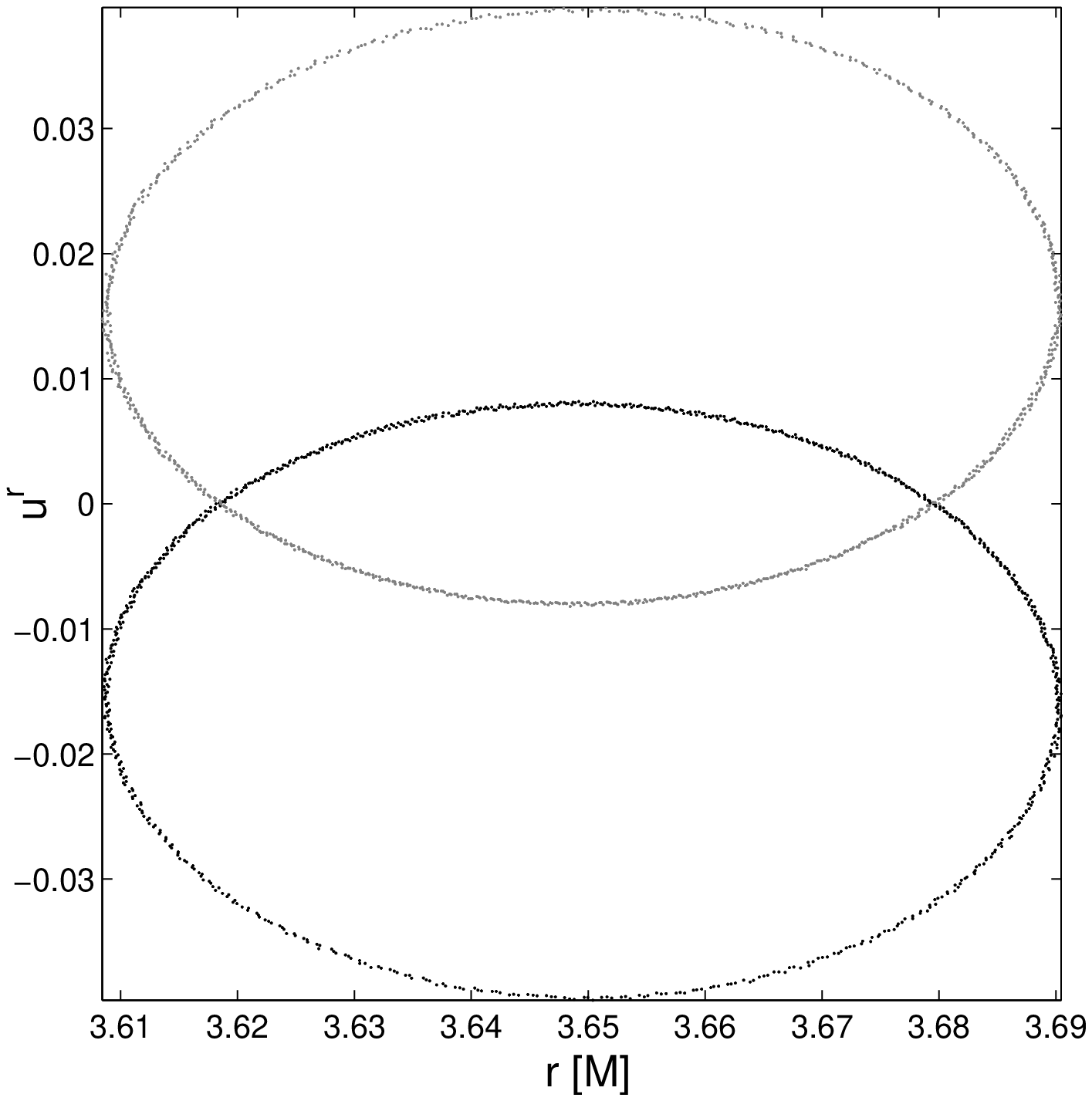}
\includegraphics[scale=.41,clip]{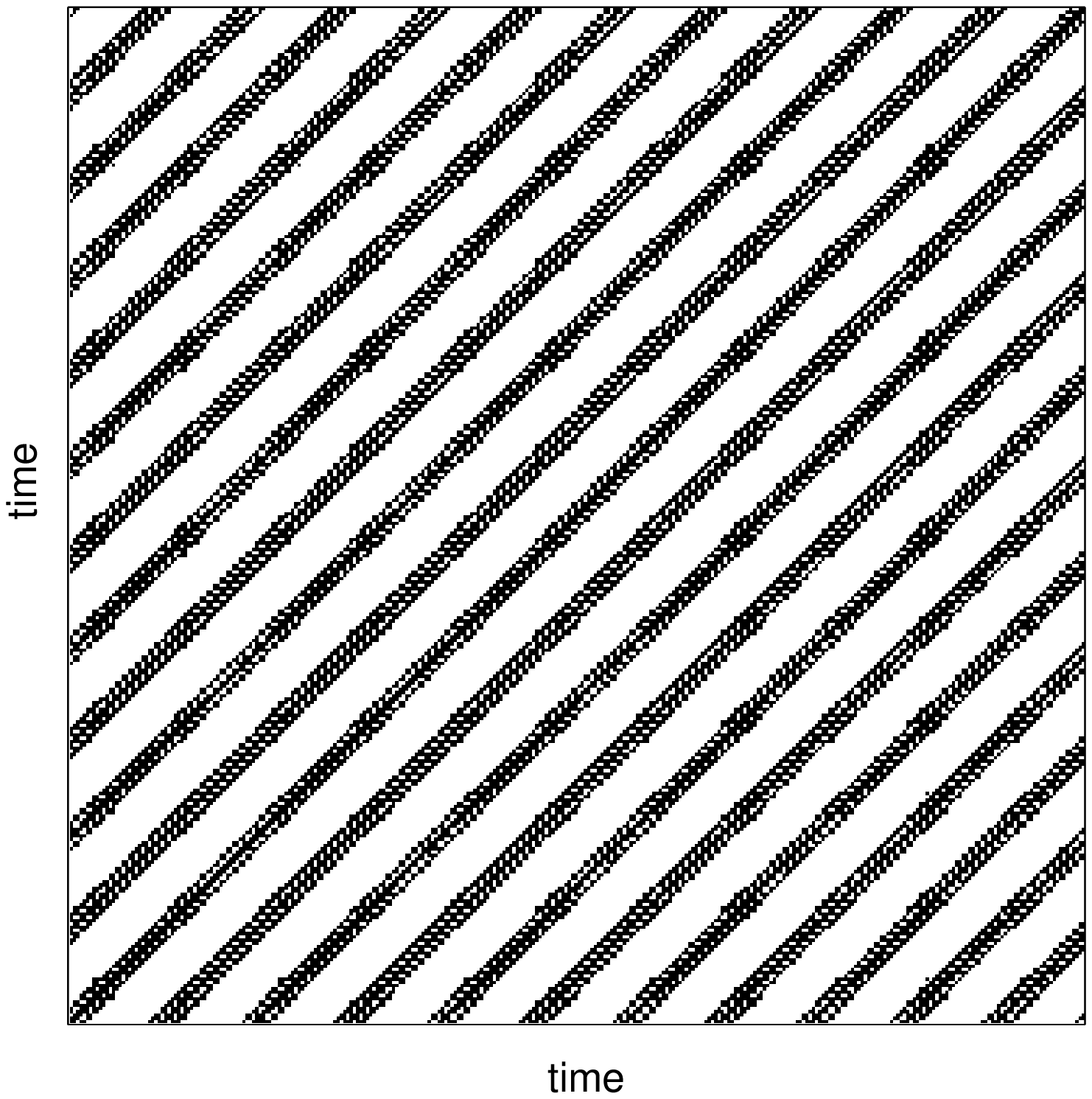}
\caption{Regular off-equatorial motion of the charged test
particle ($\tilde{E}=1.578$, $L =6M$,
$\tilde{q}\tilde{Q}=1$, $\tilde{q}B_{0}=M^{-1}$, $r(0)=3.68M$, $u^r(0)=0$ and $\theta(0)=\theta_{\rm section}=1.18$) on the Kerr background
($a=0.9\:M$) with the Wald's test field. In the Poincar\'e surface of section we distinguish $u^{\theta}\geq0$ (black point) from $u^{\theta}<0$ (grey point).
}
\end{figure}

\section{Regular and chaotic motion in off-equatorial lobes}
In this section we shall apply above described recurrence analysis
to the chosen set of orbits. From our classification of the off-equatorial
potential lobes \cite{halo2} which may appear above the horizon of the
Kerr black hole immersed in the asymptotically uniform magnetic field
(setup described by equations (\ref{kn}), (\ref{waldpot1}) and
(\ref{waldpot2})) we choose the type Id from Fig.~\ref{wald_abc} for our analysis.
In this case the off-equatorial local minima of the effective potential (\ref{effpot}) are
surrounded by the potential lobes which merge through the equatorial
plane once the energy level of the equatorial saddle point is reached
and form closed lobe which extends symmetrically above and below the
equatorial plane. Increasing the energy even more we would eventually
reach the level when this lobe breaks symmetrically in its bottom and upper parts allowing, in principle,  the particle to escape to the infinity in the axial direction. Further increase in the energy leads to the opening of the lobe towards the event horizon through the saddle point in the equatorial plane.  Described situation is captured in Fig.~\ref{traj}.

Numerical integration of the Hamilton's equations
(\ref{HamiltonsEquations}) is carried out using the multistep Adams-Bashforth-Moulton solver of variable order. In several cases when higher accuracy is demanded we employ 7-8th order Dormand-Prince method. Initial values of non-constant components of the canonical momentum $\pi_{r}(0)$ and $\pi_{\theta}(0)$ are obtained from $u^{r}(0)$ (which we set) and $u^{\theta}(0)$ which is calculated from the normalization condition $g^{\mu\nu}u_{\mu} u_{\nu}=-1$ where we always choose the non-negative root as a value of $u^{\theta}(0)$. CRP ToolBox \footnote{CRP ToolBox for Matlab: http://www.agnld.uni-potsdam.de/\~{}marwan/toolbox/} is used
for the construction of RPs and evaluation of RQA measures.

\begin{figure}[htb]
\centering\label{waldtrans}
\includegraphics[scale=0.49,clip]{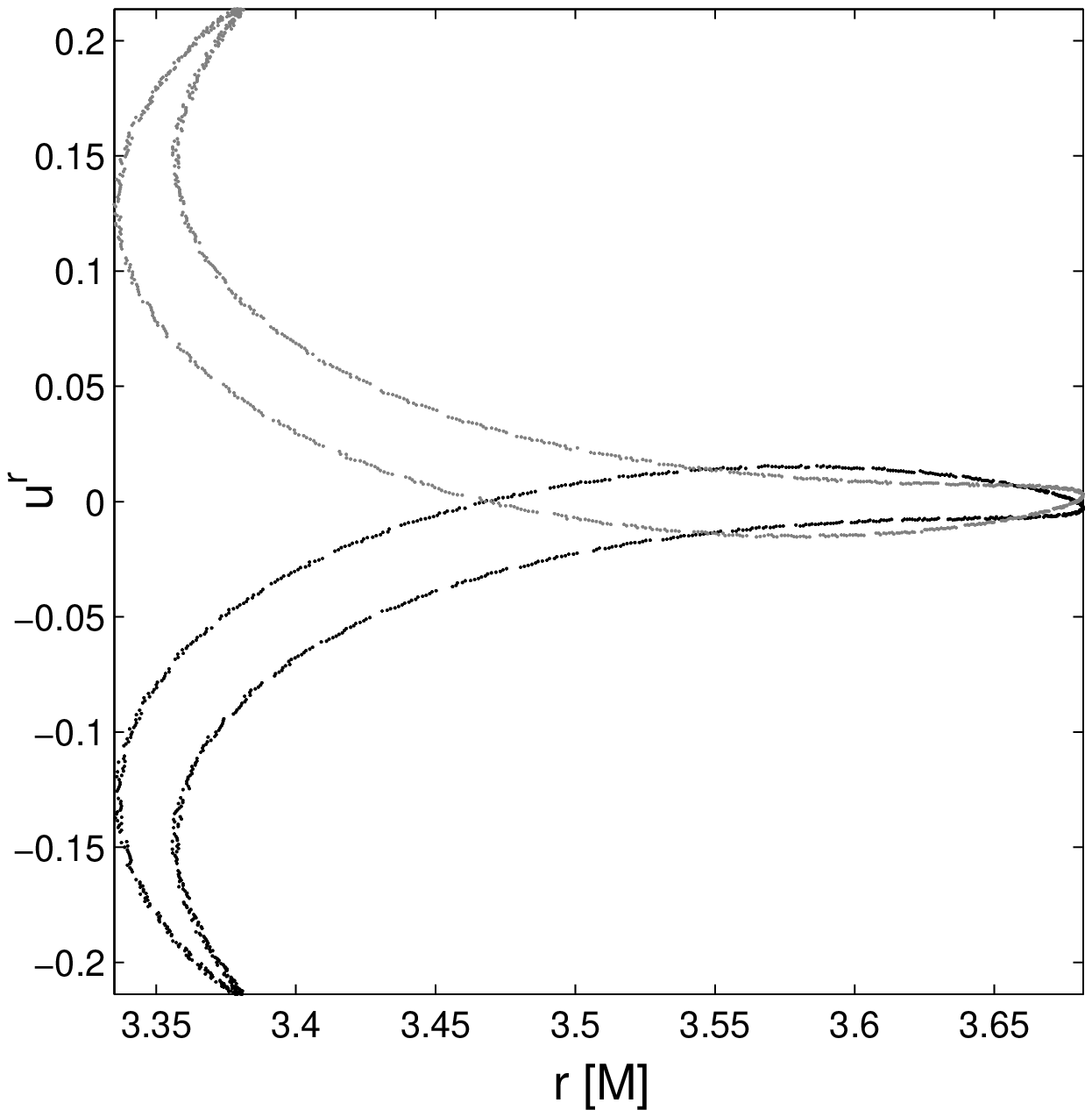}
\includegraphics[scale=0.405,clip]{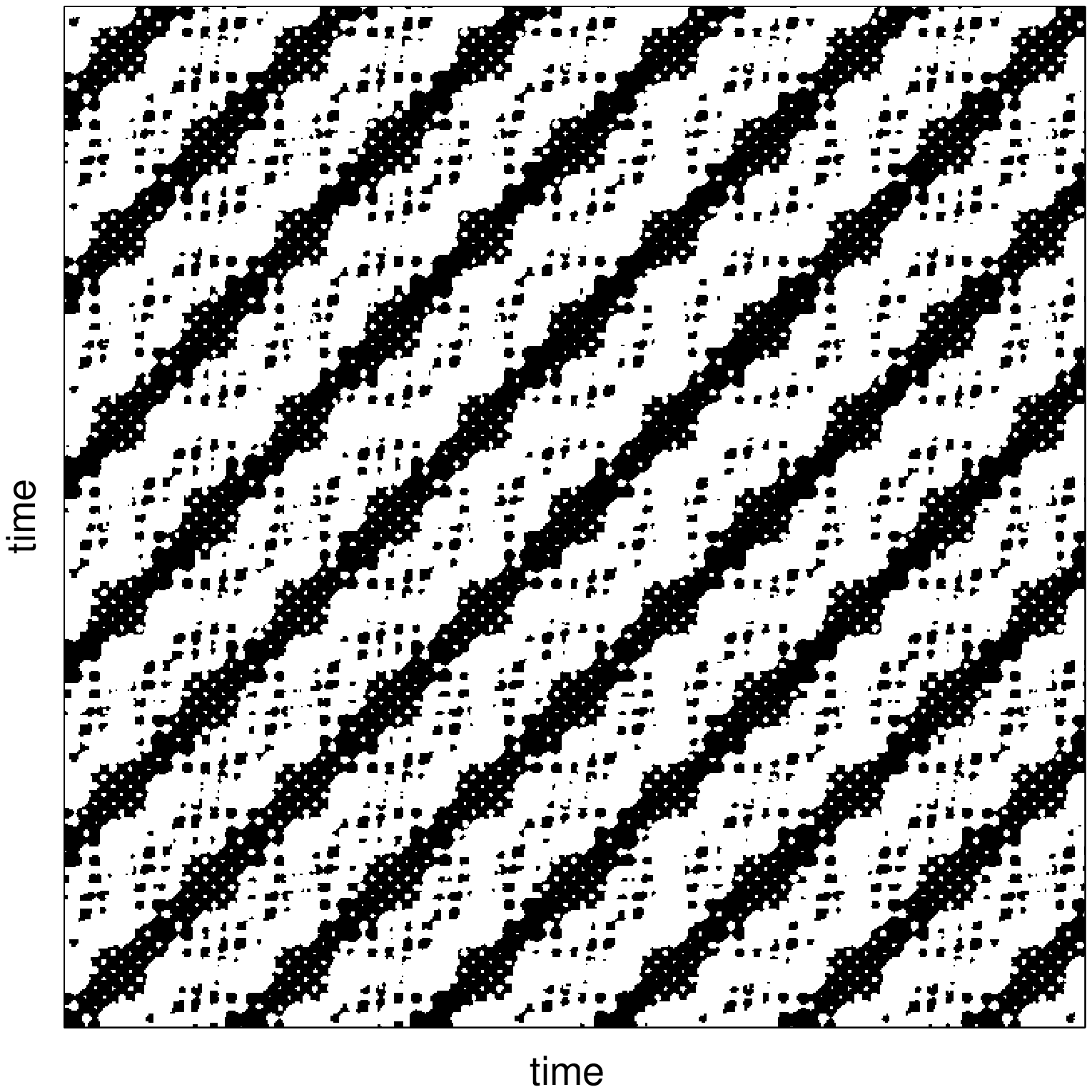}
\caption{Regular cross-equatorial motion of the charged test particle exposed to the Wald's field
which only differs from the previous case of Fig. \ref{waldreg} by
increasing energy to $\tilde{E}=1.65$ for which both off-equatorial
potential lobes are already merged via the equatorial plane.}
\end{figure}

\begin{figure}[htb]
\centering\label{waldchaos}
\includegraphics[scale=0.435,clip]{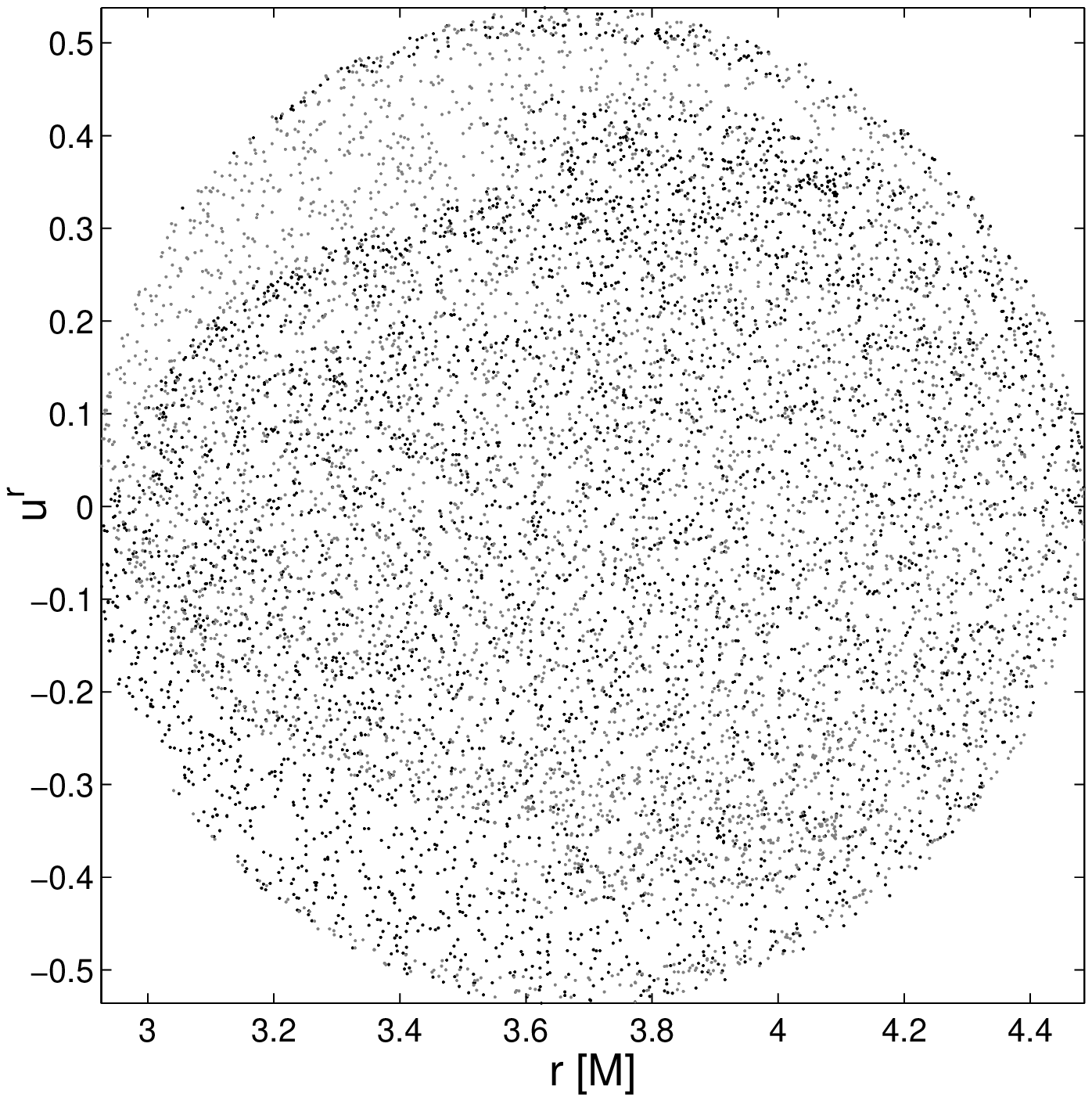}
\includegraphics[scale=0.435,clip]{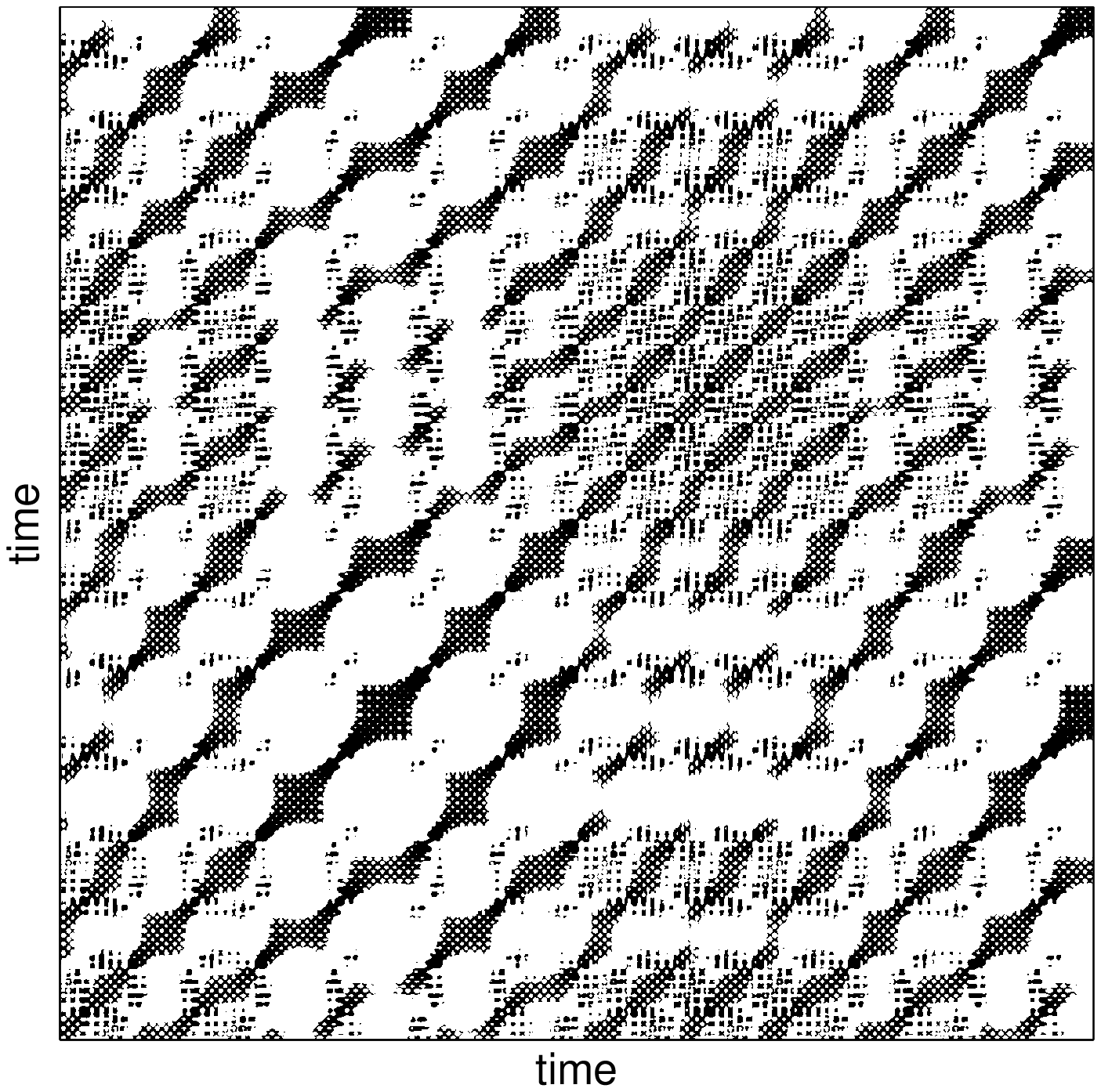}
\caption{Chaotic motion of the test particle which appears when the
energy increases to $\tilde{E}=1.75$ (while all other parameters
are kept at values of Figs. \ref{waldreg} and \ref{waldtrans}).}
\end{figure}

In Fig. \ref{kntraj} we show how does the regular trajectory in the
fully integrable system of the charged test particle in the Kerr-Newman
spacetime appear in the Poincar\'e surface of section and in the recurrence
plot. We observe that in this case the RP exhibits purely diagonal
features as expected.

Then in Fig. \ref{waldreg} we turn attention to our non-integrable
system of the charged test particle being exposed to the Wald's field on the Kerr
background. At given energy level of $\tilde{E}=1.578$ the particle
remains trapped in the off-equatorial lobe. Its motion is regular and diagonal structures in the
RP typical for the trajectories in integrable systems are preserved,
though the pattern is slightly different.

Increasing the energy to $\tilde{E}=1.65$ we obtain the lobe
already merged via the equatorial plane. To some surprise the merging
itself which occurs at $\tilde{E}\approx1.59$ is not reflected by
the change of the dynamic regime -- although
the test particle can newly cross the equatorial plane, its motion remains
regular (see Fig. \ref{waldtrans}). Nevertheless the RP proves its dynamics to be more complex than that of the integrable system of Fig. \ref{kntraj}. Diagonal structures are preserved,
though the pattern is more complicated.

However, at $\tilde{E}=1.75$ we obtain typical chaotic motion
Fig. \ref{waldchaos}. In the surface of section we observe that the
trajectory fills all allowed region (hypersurface given by the
values of integrals of motion). Diagonal lines in the RP are
seriously disrupted and complex large scale structures appear which
are characteristic indications of deterministic chaos.

\begin{figure}[htb]
\centering\label{rqa1}
\includegraphics[scale=0.8, clip]{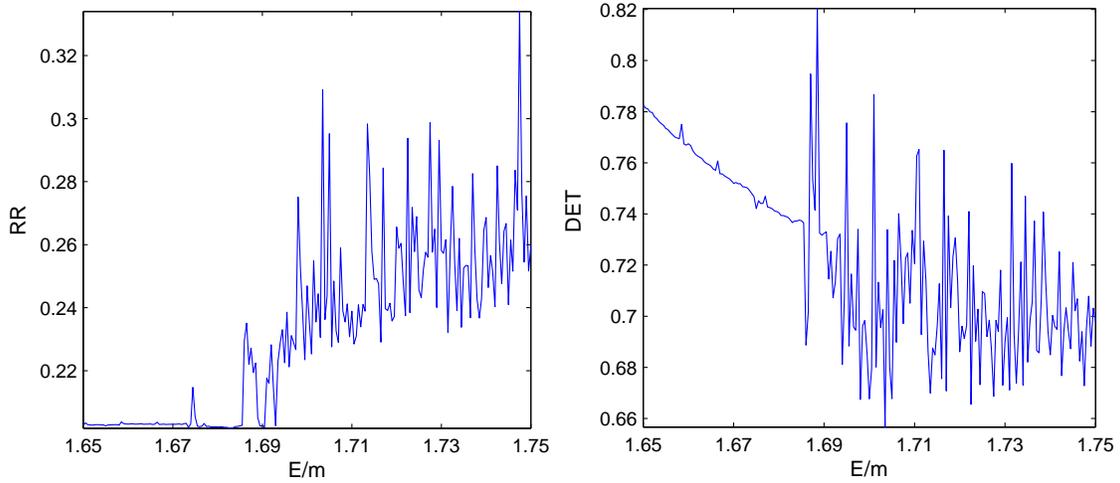}

\caption{RQA measures  $RR$ and $DET$ as a function of specific
energy $\tilde{E}$. Dramatic change of behaviour is apparent for
both quantities at $\tilde{E}\approx1.685$. This is where the chaos
sets on.}
\end{figure}

\begin{figure}[htb]
\centering
\includegraphics[scale=0.8, clip]{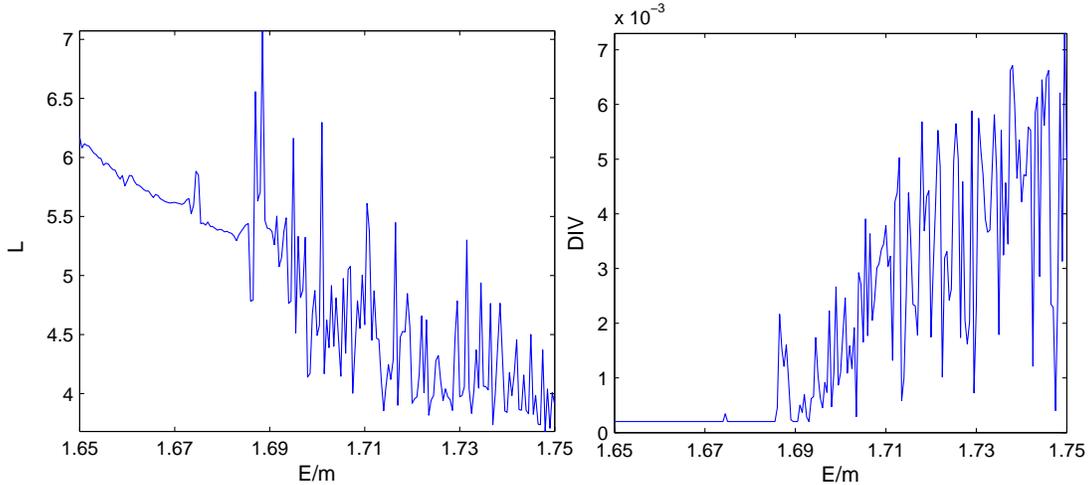}
\caption{Diagonal RQA measures  $L$ and $DIV$ as a function of
specific energy $\tilde{E}$. Apparent change of behavior at
$\tilde{E}\approx1.685$ signalizes  the onset of chaos.}
\end{figure}

\begin{figure}[htb]
\centering
\includegraphics[scale=0.8, clip]{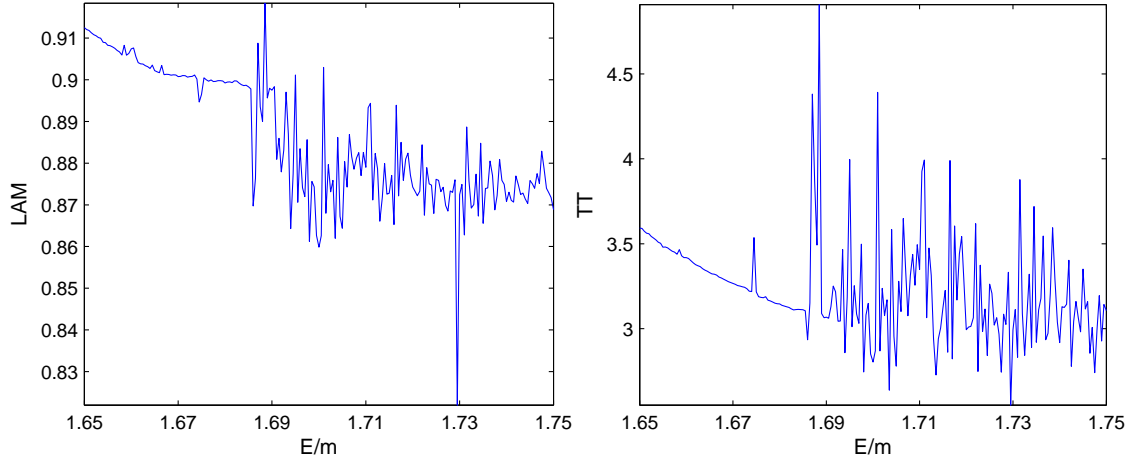}
\caption{RQA measures derived from vertical lines $LAM$ and $TT$
also clearly signalize the onset of chaos at
$\tilde{E}\approx1.685$.}
\end{figure}

\begin{figure}[htb]
\centering\label{vmax}
\includegraphics[scale=0.8, clip]{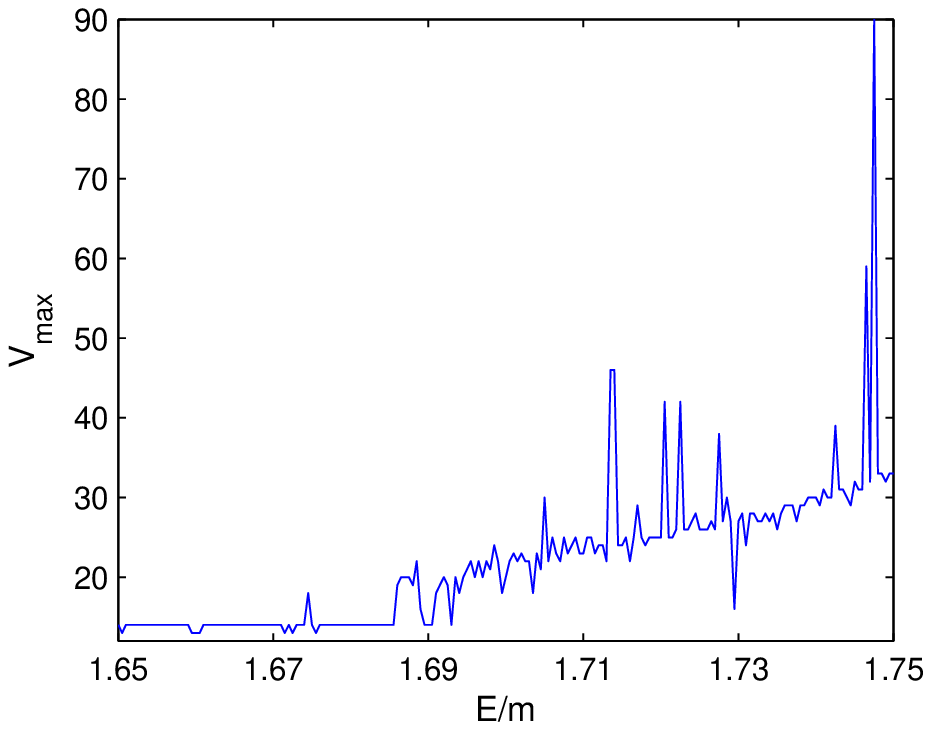}
\caption{Length of the longest vertical line  $V_{\rm{max}}$
generally rises and begins to fluctuate strongly as the chaos sets
on at $\tilde{E}\approx1.685$. }
\end{figure}

We conclude that the motion in purely off-equatorial lobes (i.e. before
merging of the lobes via equatorial plane) is regular. Surprisingly the
regularity of motion is not violated when the lobes merge and volume of the accessible portion of phase space suddenly doubles. However, increasing
further the energy eventually brings the system to the chaotic
regime of motion. This route to chaos was documented by both the
standard Poincar\'e surfaces of section and the recurrence plots. RPs
prove to act as an alternative tool to surfaces of section as they allow
to distinguish between the regular and the chaotic regime of motion by
apparent changes in their patterns.

So far we know that in our set of orbits 
the transition between the regular and the chaotic regime of the motion occurs somewhere in the range $\tilde{E}\in(1.65,1.75)$. Qualitative methods based on the visual survey of the Poincar\'e surfaces of section and the recurrence plots do not allow to localise this transition. To this end we employ recurrence quantification analysis (RQA) and observe the behaviour of RQA measures (definitions (\ref{RR})-(\ref{tt})). We calculate 200 trajectories with energies equidistantly spread over the range $\left<1.65,1.75\right>$ while other parameters of the system are fixed at values used in Figs. \ref{waldreg}--\ref{waldchaos}. In Figs. \ref{rqa1}-\ref{vmax} we observe that all queried RQA measures exhibit sudden change in its behaviour at $\tilde{E}\approx1.685$. This is where the dynamic transition between the regimes occurs. Transition from the regular motion to the chaos is thus detected not only by diagonal RQA measures but also by the vertical ones.

\section{Conclusions}
In this contribution we demonstrated that the recurrence plots and the
recurrence quantification analysis are simple, yet powerful tools
which allow one not only to decide whether the dynamic regime of
motion is regular or chaotic but also to locate (in terms of some
control parameter -- energy $\tilde{E}$ in our case) the transition between these regimes with a good precision. General relativistic context highlights the advantages of the recurrence analysis over standard methods as it operates ''locally'' (on the small length scale of $\varepsilon$) which allows for various profound computational simplifications when evaluating distances in the phase space. Complex treatment which involves evaluation of the geodesic line and measuring its length may thus be avoided.  Major drawback (cost we pay for its simplicity) of
RPs and RQA is the lack of invariance (dependence on the threshold
$\varepsilon$, $l_{\rm{min}}$,  $v_{\rm{min}}$ , choice of the norm etc.). We conclude
that RPs themselves may act as an alternative tool to Poincar\'e surfaces
of section and RQA measures are able to detect the transitions between
the dynamic regimes.

\newpage
\begin{theacknowledgments}
We thank Dr. Tom\'{a}\v{s} Pech\'{a}\v{c}ek for helpful suggestions concerning the recurrence analysis and various issues related to the nonlinear dynamical systems. Authors acknowledge support from the following grants: GAUK 119210/2010 (OK), ESA PECS98040 (VK) and GA P209/10/P190 (JK).
\end{theacknowledgments}

\bibliographystyle{aipproc}   % if natbib is available

\end{document}